\documentclass[multphys,vecphys]{svmult}

\usepackage{makeidx}     
\usepackage{graphicx}    
\usepackage{multicol}    

\makeindex             

\newcommand{\RM}[1]{\mathrm{#1}}

\newcommand{\kms}[0]{\,\RM{km}\,\RM{s}^{-1}}

\newcommand{\CII}{{C}{\sc ii}}
\newcommand{\CI}{{C}{\sc i}}
\newcommand{\OI}{{O}{\sc i}}
\newcommand{\OIII}{{O}{\sc iii}}
\newcommand{\NIII}{{N}{\sc iii}}
\newcommand{\HII}{{H}{\sc ii}}
\newcommand{\NII}{{N}{\sc ii}}
\newcommand{\msol}{M$_{\rm sun}$}
\newcommand{\cmcub}{cm$^{-3}$}

\begin{document}

\title*{CII, CI, and CO in the massive star forming region W3 Main}
\titlerunning{CII, CI, and CO in W3 Main}
\author{
Carsten Kramer\inst{1}\and
Holger Jakob\inst{1}\and
Bhaswati Mookerjea\inst{1}\and
Nicola Schneider\inst{2}\and
Martin Br\"ull\inst{1}\and
J\"urgen Stutzki\inst{1}
}
\authorrunning{Kramer et al.}
\institute{
Universit\"at zu K\"oln, I. Physikalisches Institut, Z\"ulpicher
Strasse 77, 50937 K\"oln 
\texttt{kramer@ph1.uni-koeln.de, jakob, bhaswati, bruell, stutzki}
\and 
Observatoire de Bordeaux,  Universite de Bordeaux 1, BP 89, 33270 Floirac,
France \texttt{schneider@obs.u-bordeaux1.fr}
}
%
%
\maketitle

\section{Introduction}
We used the KOSMA 3m telescope to map the core $7'\times5'$ of the
Galactic massive star forming region W3\,Main in the two fine
structure lines of atomic carbon and four mid-$J$ transitions of CO
and $^{13}$CO.  In combination with a map of singly ionized carbon
(Howe et al.  1991), and ISO/LWS data at the center position, these
data sets allow to study in detail the physical structure of the
photon dominated cloud interface regions (PDRs) where the occurance of
carbon changes from \CII\ to \CI\ to CO.
%
%
%
The physical and chemical processes in these layers are governed by
FUV radiation.  Depending on the induced chemistry, the critical
densities, and energy levels of their transitions, the lines of
[\CII], [\CI], and CO serve as probes to trace the different
temperature and density regimes.



W3\,Main is a well known site of high-mass star formation (e.g.
Tieftrunk et al. 1998).  It is associated with the W3 giant molecular
cloud complex within the Perseus arm at a distance of 2.3\,kpc
(Georgelin \& Georgelin 1976, cf.  Imai et al. 2000). W3\,Main
comprises several \HII\ regions of different sizes and evolutionary
stages, some of them hypercompact, possibly identifying the earliest
stages of high-mass star formation (Tieftrunk et al. 1997).



\section{Modelling the spatial structure of the FUV field}

Figure\,1 shows the positions of the 12 OB stars found in W3\,Main
(Tieftrunk et al. 1998).  The most luminous source is IRS\,2,
identified as O5 star. The luminosity of IRS\,5 at the map center
position corresponds to a cluster of $\sim7$ B0-type stars (Claussen
et al. 1994). We have used the spectral types of the exciting OB stars
to calculate the spatial large-scale variation of the far-UV field
heating the molecular cloud.  Figure\,1 shows the resulting FUV
distribution relative to that of [\CII] emission (Howe et al. 1991).

\begin{figure}[h]
\begin{minipage}[t]{\textwidth}
\centering
  \resizebox{9cm}{!}{
  {\includegraphics[width=4cm]{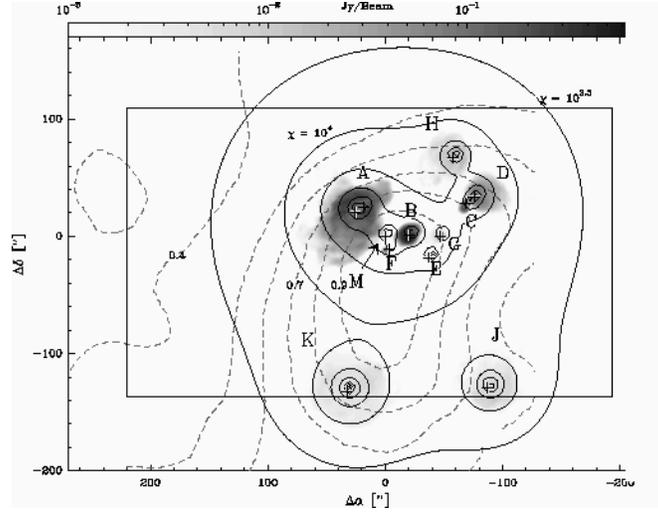}}} 
\end{minipage}
\hfill
\caption{Overlay of UV field modelled from the spectral types of 
  the OB stars known in W3\,Main, [\CII] emission, and 6\,cm radio
  continuum emission.  {\bf Grey-scale:} Image of the VLA 6\,cm radio
  continuum (Tieftrunk et al. 1997). The box denotes the region mapped
  with KOSMA in CO 3--2, 4--3, 7--6, [\CI] 1--0, 2--1, and in
  $^{13}$CO 8--7.  {\bf Solid contours:} Contours of the modelled
  spatial structure of the FUV field derived from the spectral types
  of the embedded OB stars (positions are marked with a $+$). The UV
  flux $\chi$ varies between $10^{3.5}$ at the cloud edge and $10^6$
  in the immediate vicinity of IRS\,2. Attenuation by the cloud
  density structure has yet not been taken into account. {\bf Dashed
    contours:} extended [\CII] emission observed by Howe et al. (1991)
  with the KAO at $80''$ resolution.
}
\label{fig-w3spectra}
\end{figure}

\section{IRS\,5 in W3\,Main}
\subsection{KOSMA and ISO/LWS data}

Figure\,2 shows representative spectra at the center position IRS\,5.
$^{12}$CO \newline 3--2 peak temperatures indicate an excitation temperature of
at least 50\,K.  A lower limit of the [\CI] excitation temperature of
122\,K has been derived from the [\CI] 2-1/1-0 line ratio assuming a
calibration error of 20\%.  Column densities per $75''$ beam
(0.84\,pc) and the relative abundances of the three major gas phase
species CO, CI, and CII, are derived assuming LTE. At IRS\,5, the
\CII:\CI:CO column density ratio is 18:8:74, i.e. most of the
gas-phase carbon is in CO, $\sim20$\% is in \CII, and only less than
10\% is in atomic carbon. The total H$_2$ column density derived from
$^{13}$CO 3--2 observations is $4\,10^{22}$\,cm$^{-2}$
(A$_V\sim40\,$mag) within the $75''$ beam which agrees well with the
result of Tieftrunk et al. (1998) derived from C$^{18}$O observations.

 \begin{figure}
 \begin{minipage}[t]{4cm}
   \resizebox{4cm}{!}{\includegraphics[width=4cm]{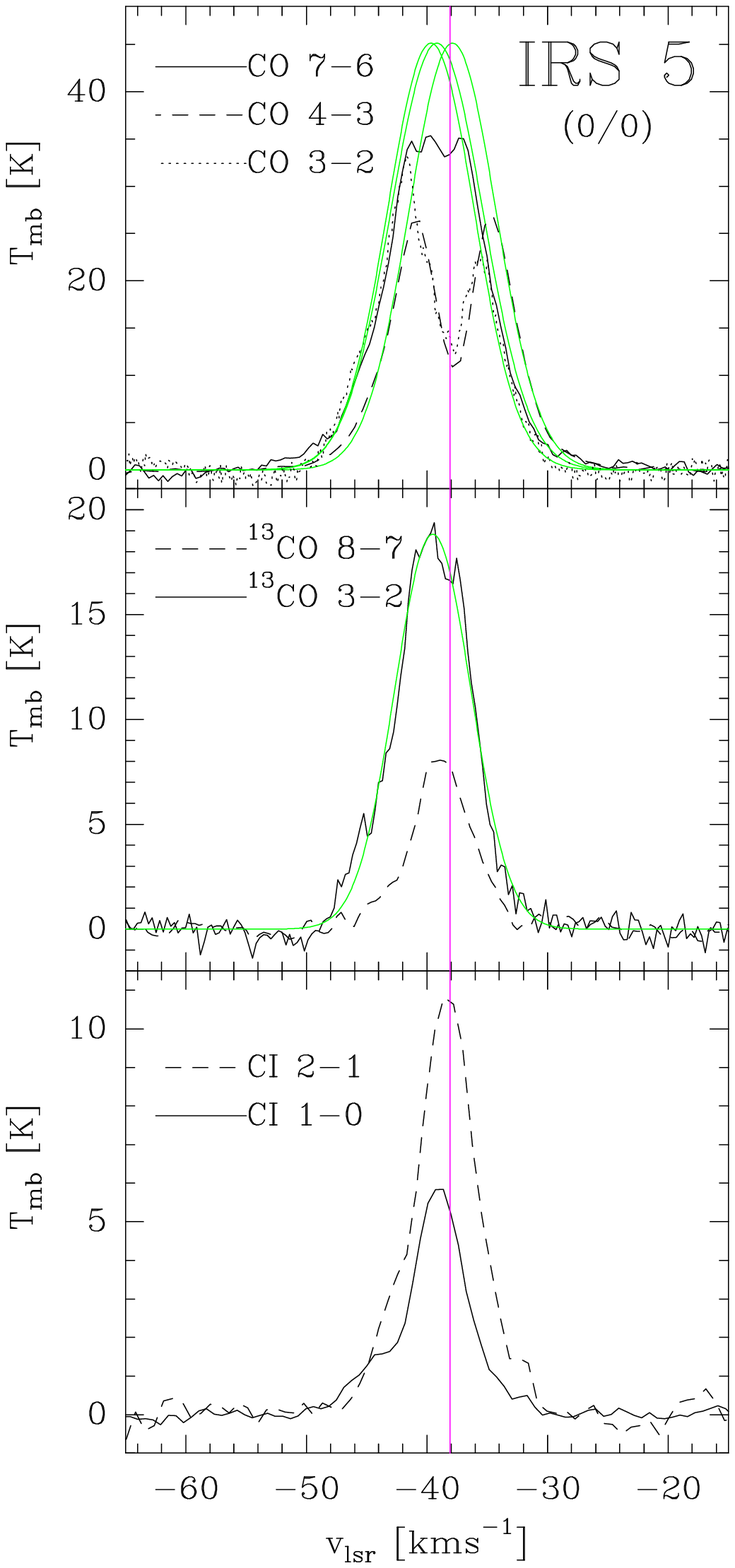}} 
 \end{minipage}
 \hspace{0.5cm}
 \begin{minipage}[t]{7cm}
 \vspace{-8.5cm}
 {\bf Figure 2:} Spectra at one representative position: IRS\,5 in
 W3\,Main.  All spectra are on a common spatial resolution of $75''$
 (0.84\,pc) and are plotted on the $T_{\rm mb}$ scale.  $^{13}$CO and
 CI lines peak at the velocity of a dip in the $^{12}$CO spectra,
 indicating that the dip is due to a colder foreground component at
 about 38\,$\kms$ (denoted by a vertical line).  To estimate the line
 integrated $^{12}$CO intensity of the background source, we fitted a
 Gaussian to the high and low velocity parts of the $^{12}$CO spectra.
 The fit results are shown.
 \end{minipage}
 \end{figure}


%

We combined the KOSMA data at IRS\,5 with ISO/LWS data from the ISO
data archive taken at the same position.  Several fine structure lines
of [\OIII], [\NIII], and [\NII] were detected tracing the ionized
medium, as well as lines of [\CII], [\OI], and four high-$J$
rotational transitions of CO tracing the neutral medium at the PDR
cloud surfaces. Figure\,3 shows the observed CO and $^{13}$CO fluxes
together with the results of a homogeneous, non-LTE radiation transfer
model. The detected line strengths indicate kinetic temperatures of
more than 140\,K and densities of more than $3\,10^4\,$\cmcub.


\begin{figure}
\begin{minipage}[t]{5cm}
  \resizebox{5cm}{!}{\includegraphics[width=5cm]{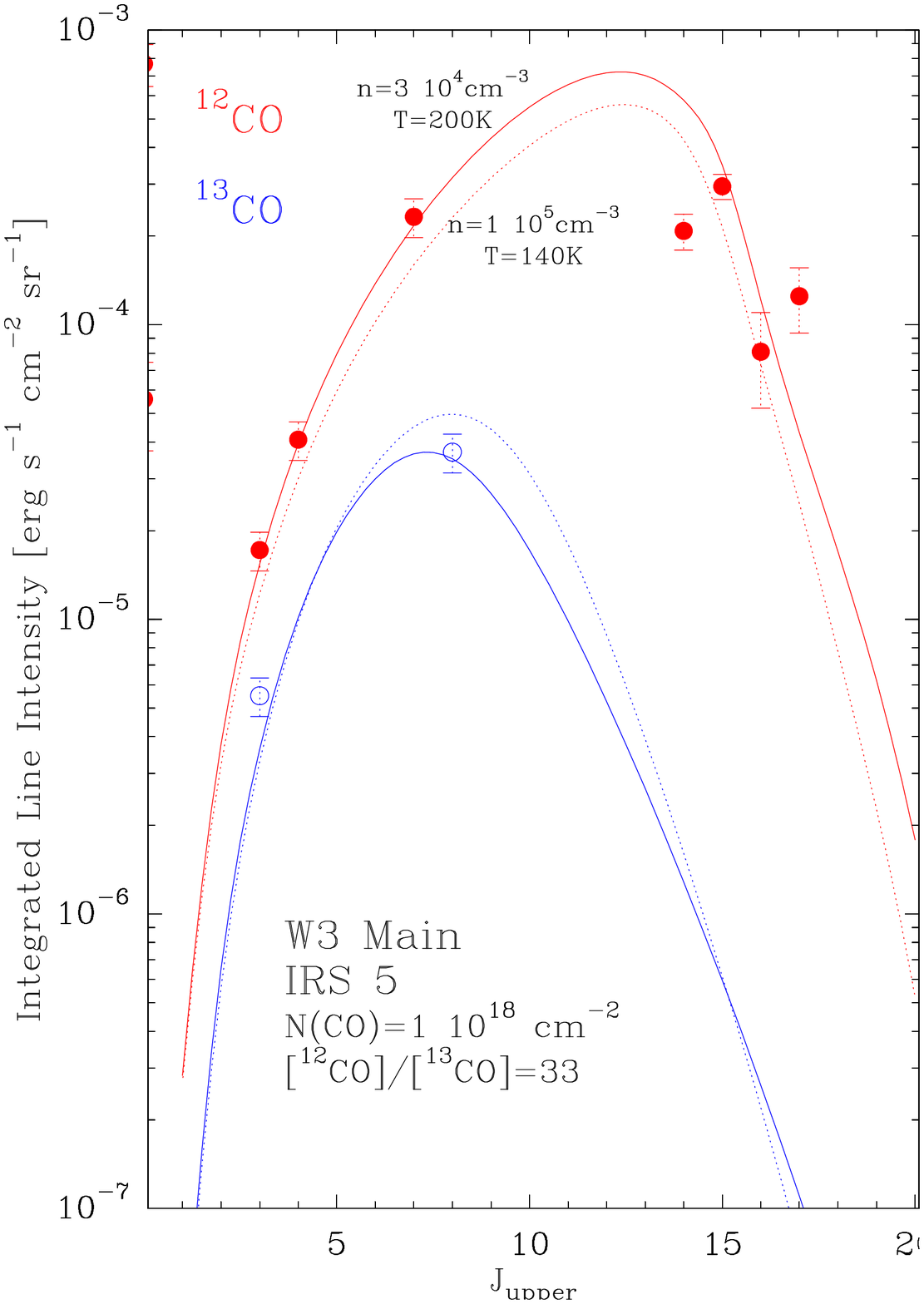}} 
\end{minipage}
\hspace{0.5cm}
\begin{minipage}[t]{6cm}
  \vspace{-6.5cm} {\bf Figure 3:} Integrated CO intensities as a
  function of upper rotational quantum number $J$ for IRS\,5 in
  W3\,Main and CO cooling curves produced by a non-LTE radiative
  transfer model.  Rotational transitions upto $J=8$ were observed
  with KOSMA, the high-$J$ transitions upto $J=17$ were
  observed with ISO/LWS.
\end{minipage}
\end{figure}

\subsection{Detailed PDR Modelling - First results}

We have started to determine the physical structure of the PDR by
detailed modelling using the FUV field previously derived. Here, we
briefly describe first results obtained from modelling the emission at
IRS\,5 using the spherical Cologne PDR model (St\"orzer et al. 1996,
2000). Its basic free parameters are the clump surface density, its
total mass, and the impinging FUV field. We kept the FUV field fixed
at $10^5$ times the Draine field as derived from the contribution of
all embedded OB stars. First results indicate a common solution for
most ratios (see Figure 4). Future modelling will include the absolute
line intensities and explore the possible importance of clumps of
different masses (clump mass spectrum), and of pre-shielding by an
embedding H$_2$ interclump medium.

\begin{figure}
\begin{minipage}[t]{\textwidth}
\centering
  \resizebox{7cm}{!}{\includegraphics[width=5cm]{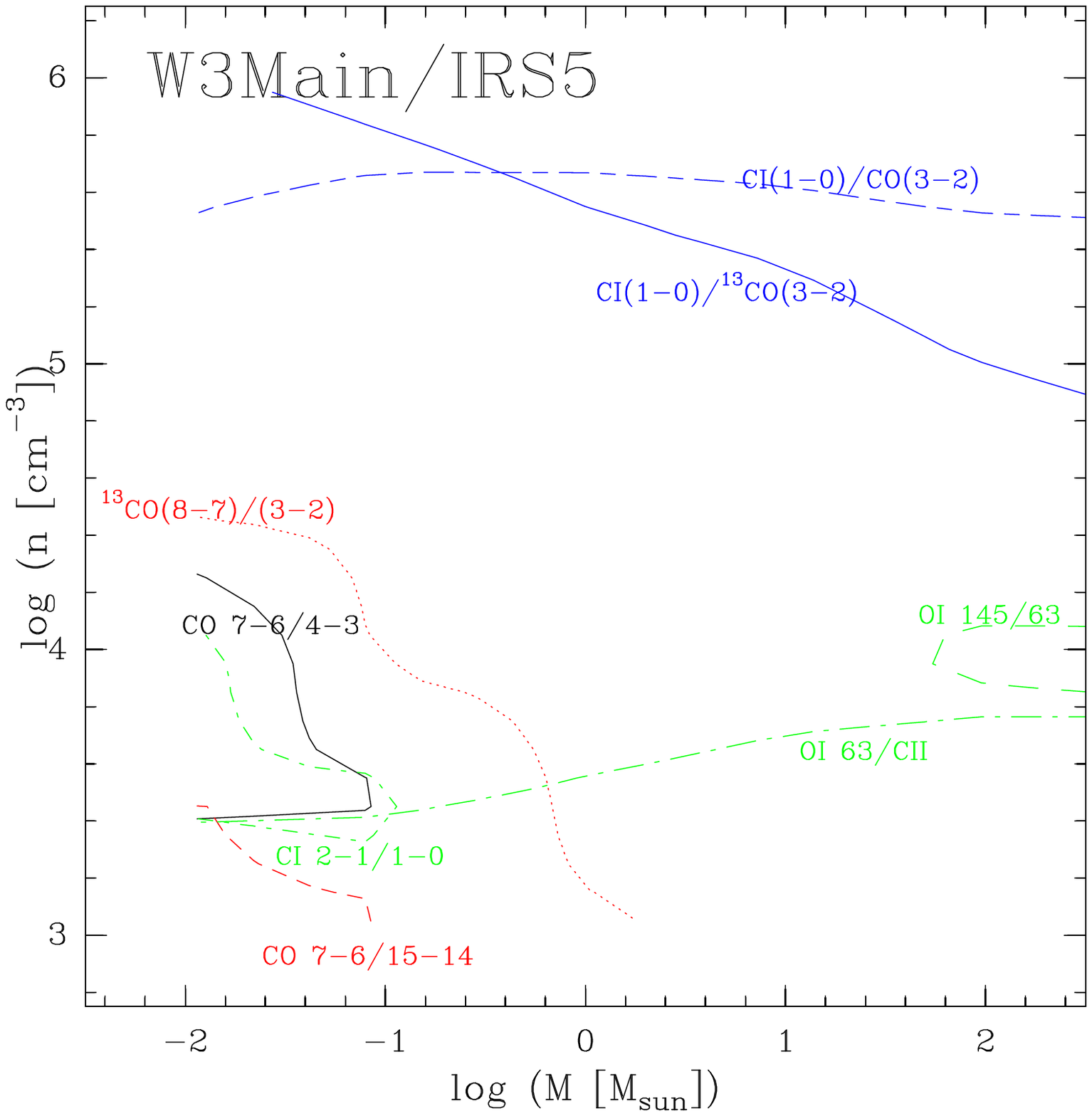}} 
\end{minipage}
\hfill
{\bf Figure 4:} Results from the PDR model for an external FUV field
of $\chi=10^5\,\chi_0$. Lines denote the observed line ratios at
IRS\,5 as a function of clump mass and clump surface density. The
[\CI]/CO and [\CI]/$^{13}$CO flux ratios indicate high volume
densities of more than $10^6$\,\cmcub. The
[\OI\,145$\mu$m/\OI\,63$\mu$m] ratio is larger than expected, probably
due to optically thick [\OI]\,63$\mu$m emission and self-absorption
due to the colder foreground material also responsible for the dips in
the low- and mid-$J$ CO lines. All other five observed ratios indicate
a common solution of about $n\approx10^4$\,\cmcub\ and a clump mass of
less than $\approx0.1$\,\msol\ in this preliminary analysis.
\end{figure}

\printindex
\end{document}